\documentclass[prd,superscriptaddress,altaffilletter,nofootinbib]{revtex4}

\usepackage{amssymb}
\usepackage[dvips]{graphicx}
\usepackage{amsmath}

\newcommand{\be}{\begin{equation}}
\newcommand{\ee}{\end{equation}}
\newcommand{\bea}{\begin{eqnarray}}
\newcommand{\eea}{\end{eqnarray}}

\newcommand{\der}{\partial}
\newcommand{\vphi}{\varphi}

\begin{document}

\title{An issue with the classification of the scalar-tensor theories of gravity}

\author{Israel Quiros}\email{iquiros@fisica.ugto.mx}\affiliation{Dpto. Ingenier\'ia Civil, Divisi\'on de Ingenier\'ia, Universidad de Guanajuato, Gto., M\'exico.}

\author{Roberto De Arcia}\email{robertodearcia@gmail.com}\affiliation{Dpto. Ingenier\'ia Civil, Divisi\'on de Ingenier\'ia, Universidad de Guanajuato, Gto., M\'exico.}

\author{Ricardo Garc\'{\i}a-Salcedo}\email{rigarcias@ipn.mx}\affiliation{Instituto Polit\'ecnico Nacional, CICATA-Legaria, Ciudad de M\'exico, 11500, M\'exico.}

\author{Tame Gonzalez}\email{tamegc72@gmail.com}\affiliation{Dpto. Ingenier\'ia Civil, Divisi\'on de Ingenier\'ia, Universidad de Guanajuato, Gto., M\'exico.}

\author{Francisco Antonio Horta-Rangel}\email{anthort@hotmail.com}\affiliation{Dpto. Ingenier\'ia Civil, Divisi\'on de Ingenier\'ia, Universidad de Guanajuato, Gto., M\'exico.}

\date{\today}

\begin{abstract} In the bibliography a certain confusion arises in what regards to the classification of the gravitational theories into scalar-tensor theories and general relativity with a scalar field either minimally or non-minimally coupled to matter. Higher-derivatives Horndeski and beyond Horndeski theories that at first sight do not look like scalar-tensor theories only add to the confusion. To further complicate things, the discussion on the physical equivalence of the different conformal frames in which a given scalar-tensor theory may be formulated, makes even harder to achieve a correct classification. In this paper we propose a specific criterion for an unambiguous identification of scalar-tensor theories and discuss its impact on the conformal transformations issue. The present discussion carries not only pedagogical but also scientific interest since an incorrect classification of a given theory as a scalar-tensor theory of gravity may lead to conceptual issues and to the consequent misunderstanding of its physical implications.\end{abstract}


\maketitle


\section{Introduction}

Due to the yet unknown nature of the dark components of the cosmic background alternatives to the general theory of relativity like the Brans-Dicke (BD) theory, scalar-tensor theories (STT) of gravity and their higher derivative generalizations -- collectively known as Horndeski \cite{horndeski_gal, nicolis_gal, deffayet_vikman_gal, deffayet_deser_gal, deffayet_prd_2011, kazuya_gal, fab_4_prl_2012, clifton-phys-rept-2012, deffayet-rev, tsujikawa_lect_not, kazuya_rpp_2016, quiros-ijmpd-rev, kobayashi-rpp-rev, odi-1, nojiri-rev, kase} and beyond Horndeski \cite{kobayashi-rpp-rev, bhorn-langlois, bhorn-fasiello, bhorn-crisostomi, bhorn(vainsh), mancarella-jcap-2017, bhorn-ostrog, chagoya-tasinato-jhep-2017} theories -- have acquired renewed interest. The use of scalar fields in the gravitational theories has a theoretically-motivated origin in the famous theorem by Lovelock \cite{lovelock}. According to this theorem, in four dimensions the unique metric higher-derivative theory that gives rise to second-order motion equations, is based in the Einstein-Hilbert action. The resulting motion equations are just the Einstein's equations of general relativity (GR). As a consequence of the Lovelock's theorem, if one wants to construct metric theories of gravity with field equations differing from those of GR, one is left with a few options \cite{clifton-phys-rept-2012}: Either i) accept higher derivatives of the metric in the field equations, or ii) adopt higer-dimensional spacetimes, or iii) consider other fields beyond the metric tensor, among other more exotic possibilities.

In the Brans-Dicke theory \cite{bd-1961} the scalar field plays the role of a point-dependent gravitational coupling that sets the strength of the gravitational interactions point by point. Hence, the carriers of the gravitational interactions are the two polarizations of the graviton and the scalar field. Scalar-tensor theories of gravity \cite{quiros-ijmpd-rev, stt-s} are a generalization of the BD theory where the coupling parameter is a function of the scalar field: $\omega_\text{BD}\rightarrow\omega(\phi)$, i. e., it is varying parameter. The need for a generalization of BD theory comes from the tight constraints on $\omega_\text{BD}$ that the solar system experiments have established \cite{will-lrr-2014}. If $\omega_\text{BD}$ were a varying coupling then the latter experimental constraints may be avoided or, at least, alleviated. The Horndeski theories represent further extension of scalar-tensor theories by including higher derivatives of the scalar field while keeping the dynamics governed by second-order motion equations \cite{horndeski_gal, nicolis_gal, deffayet_vikman_gal, deffayet_prd_2011, fab_4_prl_2012, deffayet-rev, tsujikawa_lect_not, kazuya_rpp_2016}. Hence, the Horndeski constructions avoid the Ostrograsky instability \cite{ostrog, woodard}. Unlike this the beyond Horndeski theories are characterized by equations of motion with 3rd order derivatives although only three degrees of freedom (d.o.f) are propagated \cite{bhorn-langlois}, i. e., the gravitational interactions are mediated by 3 d.o.f like in the Horndeski theories.

No matter how trivial the differences between GR and the STT may look, one finds several examples in the bibliography where an incorrect identification of a theory as belonging in the STT class is made. A clear example of the latter statement is the so called Saez-Ballester theory \cite{saez-ballester}. It is given by the following action (in this paper we use the units system with $8\pi G_N=c^2=\hbar=1$): 

\bea S_\text{SB}=\frac{1}{2}\int d^4x\sqrt{|g|}\left[R-\omega\vphi^r(\der\vphi)^2\right],\label{saez-ballester-action}\eea where $(\der\vphi)^2\equiv g^{\mu\nu}\nabla_\mu\vphi\nabla_\nu\vphi$ and $\omega$, $r$ are free constants. Under the innocuous redefinition of the scalar field; $$\phi=\frac{2\sqrt\omega}{r+2}\,\vphi^\frac{r+2}{2},$$ the above action is transformed into general relativity with a scalar field as matter source: $$S=\frac{1}{2}\int d^4x\sqrt{|g|}\left[R-(\der\phi)^2\right].$$ Hence, any solution the Saez-Ballester theory is supposed to give to an issue arising in the framework of GR, may be necessarily spurious. Although this seems like a trivial conclusion, lots of work published on the issue incorrectly identify the above theory as a modification of general relativity (as a scalar-tensor theory to be precise). For a non-exhaustive sample see references \cite{socorro-2010, jamil-epjc-2012, sharif-prd-2014, rao-2017, aditya-2018, rao-2018, mishra-2019}. 

An additional source of confusion may be related with the issue on the physical equivalence between the different conformal frames in which a given scalar-tensor theory can be formulated, also known as the 'conformal transformations issue' \cite{dicke-1962, faraoni, faraoni_prd_2007, sarkar_mpla_2007, sotiriou, deruelle, quiros_grg_2013}. According to several authors \cite{faraoni_prd_2007, deruelle} a given STT is physically equivalent to GR with a scalar field that is non-minimally coupled to the matter degrees of freedom. If this point of view were correct then there would not be physical distinction between GR with an additional universal fifth-force and the STT, i. e., between metric and non-metric theories of gravity. This would make even harder to achieve a correct unambiguous classification of the STT-s. Additional uncertainty may arise due to the presence of higher derivatives of the scalar field and to complicated self-couplings.

In this paper we shall discuss on these issues on the basis of a unambiguous criterion of what to regard as a scalar-tensor theory. The issue has been put forward in the review \cite{quiros-ijmpd-rev} where in subsection 7.1.1 a related brief discussion can be found. However, we feel that the discussion in the mentioned reference was not as exhaustive as required given to the subtlety of the issue. Aim of the present paper is, precisely, to give new arguments and to improve the discussion on the classification of the STT-s. The proposed criterion is the effective gravitational coupling, i. e., the one that is measured in Cavendish-type experiments. It is usually computed in the weak-field limit of the theory and also by means of the cosmological perturbations approach. In the simplest known situations it can be written as a product of the inverse factor that multiplies the curvature scalar in the action -- a function of the scalar field $f^{-1}(\phi)$ -- by a function $h$ of the coupling parameter $\omega$: $$8\pi G_\text{eff}=f^{-1}(\phi)h(\omega),$$ where the left-hand factor $f^{-1}$ is due to the tensor contribution to the gravitational interactions, while the factor $h$ is originated from the scalar field's contribution to the gravitational effects. However, in more complex cases when higher-order derivatives and self-couplings of the scalar field are considered, the effective coupling is a rather non-trivial quantity (see subsection \ref{sect-horn-geff} below). Whenever this effective gravitational coupling $G_\text{eff}$ is a function of the scalar field and/or of its derivatives and, also, of the curvature of spacetime, the given theory of gravity may be regarded as a STT. Otherwise, if the effective coupling is a constant, the resulting theory is indistinguishable from general relativity. 

That the proposed criterion for a unambiguous classification of STT-s is not as trivial as it seems, is clear from the following example. Let us choose, for instance, the gravitational coupling itself, i. e., the function of the scalar field that multiplies the curvature scalar in the action, as a measure for differentiating the STT-s from GR. There are theories in the Horndeski class for which the gravitational coupling is a constant, and such that the derivative couplings are with the scalar field itself (and with its derivatives) and not with the curvature. One example of such a theory is the cubic galileon, whose action may be written in the following form: $$S=\int d^4x\sqrt{|g|}\left[\frac{1}{2}R+X-V-G_3\Box\phi\right],$$ where $X\equiv-(\der\phi)^2/2$ is the kinetic energy density of the scalar field, $\Box\equiv g^{\mu\nu}\nabla_\mu\nabla_\nu$, and $G_3=G_3(\phi,X)$. From this action, if follow the classification based on the gravitational coupling (in the present example it is just $1/2$ due to our choice of units), it is apparent that the theory belongs in the GR class. On the contrary, if follow the classification determined by the effective gravitational coupling, as we shall show in section \ref{sect-horn-sst}, this is, in fact, a scalar-tensor theory. Just to underline the subtlety in this case let us anticipate that if $G_3=G_3(\phi)$ were a function of the scalar field exclusively, the theory were indistinguishable from GR according to either classification. Otherwise, for the above theory to be a STT, it is required that $G_3=G_3(X)$. Hence, the higher-order derivative contribution is what makes of this theory an STT.

In what regards to the additional ambiguity in the classification of the STT-s in connection with the conformal transformations' issue, the situation may be briefly stated in the following way. Under a conformal transformation of the metric:\footnote{As discussed in \cite{fulton_rmp_1962}, conformal transformations can be formulated in different ways so that it is very important to distinguish these different formulations because they have different physical interpretation and because equations belonging to different formulations may easily be confused. Such a confusion can lead to mathematical inconsistencies even when two formulations are equivalent. Accordingly, in the present paper in order to be specific, we assume that the conformal transformation of the metric does not represent a diffeomorphism or, properly, a conformal isometry (see \cite{dicke-1962} or the appendix D of Ref. \cite{wald_book_1984}). Moreover, the spacetime points -- same as spacetime coincidences or events -- as well as the spacetime coordinates that label the points in spacetime, are not modified or altered by the conformal transformations in any way. }

\bea g_{\mu\nu}\rightarrow\Omega^{-2}g_{\mu\nu}\;\left(g^{\mu\nu}\rightarrow\Omega^2g^{\mu\nu}\right),\label{conf-t}\eea where $\Omega^2=\Omega^2(x)$ is the (non-vanishing) conformal factor, a given scalar-tensor theory can be transformed into general relativity with an additional universal fifth-force acting on the matter degrees of freedom with non-vanishing mass. Hence, if we assume that the different conformal frames in which a given STT may be formulated are physically equivalent, there may not exist distinction between non-metric and metric theories of gravity. If assume, on the contrary, that the different conformal frames are not physically equivalent and that there exists a physical metric among the conformally related metrics, the following ambiguity arises. It follows from the low-energy limit of string theory \cite{wands-rev} that the Lagrangian of the different matter degrees of freedom $\psi^{(i)}$ can be written as: ${\cal L}_m(\psi^{(i)},g^{(i)}_{\mu\nu})$, where it is explicit that the different matter species $\psi^{(i)}$ couple to different conformal metrics $g^{(i)}_{\mu\nu}$;

\bea g^{(i)}_{\mu\nu}=e^{2\beta_i\phi}g_{\mu\nu},\label{cham-coup}\eea where $\phi$ is the dilaton (the scalar field), $g_{\mu\nu}$ is the Einstein's frame (EF) metric and the $\beta_i$-s are dimensionless constants of order unity \cite{cham-khoury}. Given that the different matter species follow geodesics of different conformal metrics, the question then is: which one of the different conformal metrics is the physical one? Obviously there would not be a satisfactory answer to the above question.

We want to underline that the scope of the present paper is limited to standard (torsionless) scalar-tensor theories and to their generalizations that belong in the class of the Horndeski theories. This means that such modifications of the gravitational theory as the teleparallel theories \cite{saridakis-2011, saridakis-rpp-2016}, the 'beyond Horndeski' theories \cite{kobayashi-rpp-rev, bhorn-langlois, bhorn-fasiello, bhorn-crisostomi, bhorn(vainsh), mancarella-jcap-2017, bhorn-ostrog, chagoya-tasinato-jhep-2017}, the so called new gravitational scalar-tensor theories \cite{naruko-2016, saridakis-2016}, where the Lagrangian density depends not only on the Ricci scalar but also on its first and second derivatives: ${\cal L}={\cal L}\left[R,(\der R)^2,\Box R\right]$, and the extended theories of gravity \cite{chiba-jcap-2005, capozziello-grg-rev, capoz-phys-rept-rev, gottlober-cqg-1990, schmidt-cqg-1990, wands-cqg-1994, capoz-etg-grg-2000, nojiri-ijgmp-2007, capoz-etg-prd-2015} whose Lagrangian density depends on higher-order derivatives of the Ricci scalar ${\cal L}={\cal L}\left(R,\Box R,\Box^2R,...,\Box^k R,\phi\right)$, are excluded from the present discussion. These modifications require of separate consideration.

The paper has been organized in the following way. The basic elements of Brans-Dicke and scalar-tensor theories are exposed in section \ref{sect-bd}, while the fundamentals of Horndeski theories are given in section \ref{sect-horn-theor}. In particular the subsection \ref{sect-horn-geff} is dedicated to briefly expose the derivation of the expression for the effective gravitational coupling, i. e., the one measured in Cavendish-type experiments. Both the well-known derivation based on the post-Newtonian approximation and the less-known derivation that is based on cosmological perturbations of the background, are exposed. Sections \ref{sect-bd} and \ref{sect-horn-theor} have been included in order for the paper to be self-contained but the reader which is familiar with these topics may safely skip them. A detailed discussion on what to consider as an adequate measure for the classification of theories into STT or GR, is given in section \ref{sect-horn-sst}. In section \ref{sect-fdr} we briefly discuss on this issue on the light of the mathematical equivalence between $f(R)$ and Brans-Dicke theories. The impact of the conformal transformations' issue on the classification is discussed in section \ref{sect-ct}, while in section \ref{sect-discu} we give some final remarks on the classification of scalar-tensor gravitational theories. Brief conclusions are given in section \ref{sect-conclu} where the main achievements of the paper are summarized.


\section{Brans-Dicke and scalar-tensor theories of gravity}\label{sect-bd}

The Brans-Dicke (BD) theory of gravity \cite{bd-1961} is the prototype of STT. It is thought to embody the Mach's principle \cite{dicke_am_sci_1959, rothman_am_sci_2017}. Mathematically the BD theory is expressed by the following action principle:\footnote{For the physical principles which the BD theory is based on we recommend Ref. \cite{fujii_book_2004}.}

\bea S=\int d^4x\sqrt{|g|}\left[\phi R-\frac{\omega_\text{BD}}{\phi}(\der\phi)^2-2V(\phi)+2{\cal L}_m(\psi^{(i)},\nabla\psi^{(i)},g_{\mu\nu})\right],\label{bd-action}\eea where ${\cal L}_m$ is the Lagrangian of the matter fields $\psi^{(i)}$, $\phi$ is the BD scalar field, $V(\phi)$ is the self-interaction potential for $\phi$, and $\omega_\text{BD}$ is a free constant -- the only free parameter of the theory -- called as the BD parameter. It should be noticed that in the original formulation of the BD theory \cite{bd-1961} the scalar field's self-interaction potential was not considered, a case usually called as massless BD theory.\footnote{The action \eqref{bd-action}, with perhaps a quite different aspect and by ignoring the scalar field's self-interaction term, was first given by other scientists including Pascual Jordan \cite{jordan, brans_phd_thesis, goenner_grg_2012}, this is why the BD theory is sometimes called as Jordan-Brans-Dicke (JBD) theory. For a nice historical account of the development of the JBD theory we recommend Ref. \cite{goenner_grg_2012}.} In the form depicted by the action \eqref{bd-action}, the BD theory is said to be given in the Jordan frame (JF). In the JFBD theory \eqref{bd-action} the scalar field plays the role of the point-dependent gravitational coupling (not the same as the measured Newton's constant as shown below): 

\bea \phi=\frac{1}{8\pi G(x)}=M^2_\text{Pl}(x),\label{grav-coup}\eea where $M_\text{Pl}(x)$ is the point-dependent reduced Planck mass. The BD scalar field sets the strength of the gravitational interactions at each point in spacetime. In consequence, this is not a completely geometrical theory of gravity since the gravitational effects are encoded not only in the curvature of the spacetime but, also, in the interaction with the propagating scalar field degree of freedom.

From \eqref{bd-action}, by varying with respect to the metric, the Einstein-Brans-Dicke (EBD) equations of motion can be derived (see Ref. \cite{ejp-2016} for the details of the derivation):

\bea G_{\mu\nu}=\frac{1}{\phi}\left[T^{(\phi)}_{\mu\nu}+T^{(m)}_{\mu\nu}\right]+\frac{1}{\phi}(\nabla_\mu\nabla_\nu-g_{\mu\nu}\Box)\phi,\label{bd-feq}\eea where $G_{\mu\nu}\equiv R_{\mu\nu}-g_{\mu\nu}R/2$ is the Einstein's tensor,

\bea T^{(\phi)}_{\mu\nu}\equiv\frac{\omega_\text{BD}}{\phi}\left[\der_\mu\phi\der_\nu\phi-\frac{1}{2}g_{\mu\nu}\left(\der\phi\right)^2\right]-g_{\mu\nu}V(\phi),\label{def-1}\eea is the stress-energy tensor of the BD-field, and $T^{(m)}_{\mu\nu}$ is the stress-energy tensor of the matter degrees of freedom. By taking variations of \eqref{bd-action} with respect to the BD field, the following ``Klein-Gordon-Brans-Dicke'' (KGBD) equation of motion is obtained (see Ref. \cite{ejp-2016} for details):

\bea (3+2\omega_\text{BD})\Box\phi=2\phi\der_\phi V-4V+T^{(m)}.\label{kgbd-eq}\eea Besides, the standard conservation equation for the stress-energy tensor of the matter fields is obtained: $\nabla^\nu T^{(m)}_{\nu\mu}=0$. This entails that the matter fields respond only to the metric $g_{\mu\nu}$, i. e., these follow geodesics of that metric. Hence, what is the role of the scalar field in the gravitational interactions of matter? As seen from equations \eqref{bd-feq} and \eqref{kgbd-eq} above, the matter acts as a source of the metric and of the scalar fields and, then the metric says back the matter how it should move. The scalar field just modulates the strength of the interactions of matter with the metric field through the gravitational coupling.  

As said the gravitational coupling \eqref{grav-coup} is not the one measured in Cavendish-like experiments. The effective gravitational 'constant' , $G_\text{eff}$, i. e., the one that is really measured, can be found in the weak-field limit of the theory. It is obtained that \cite{bd-1961, fujii_book_2004, brans_phd_thesis, weak-f-bd-2}:

\bea 8\pi G_\text{eff}=\frac{1}{\phi_0}\left(\frac{4+2\omega_\text{BD}}{3+2\omega_\text{BD}}\right),\label{bd-eff-g}\eea where the scalar field is evaluated today, $\phi_0=\phi(t_0)$, and it is determined by appropriate cosmological boundary conditions given far from the system of interest. This means that the factor $\phi\approx \phi_0$ that multiplies the curvature scalar in \eqref{bd-action} -- the one that sets the strength of the gravitational interactions point by point -- is just the gravitational coupling associated with the tensor part of the gravitational interaction. Meanwhile, the effective gravitational coupling constant, $G_\text{eff}$, is also contributed by the scalar piece of the gravitational interactions, the one that originates the strange factor $(4+2\omega_\text{BD})/(3+2\omega_\text{BD})$ in \eqref{bd-eff-g}. 

From equation \eqref{bd-eff-g}, it is evident how the GR limit can be recovered from the BD theory: just take the $\omega_\text{BD}\rightarrow\infty$ limit. In this limit it follows that $8\pi G_\text{eff}=\phi_0^{-1}$, while for a stationary mass point of mass M we get that: $$g_{00}=-1+\frac{2G_\text{eff} M}{r},\;g_{ij}=\delta_{ij}\left(1+\frac{2G_\text{eff} M}{r}\right).$$ The fact that in the (weak-field) $\omega_\text{BD}\rightarrow\infty$ limit the measured gravitational constant $8\pi G_\text{eff}=1/\phi_0$, means that the strength of the gravitational interaction in this limit is entirely due to the metric tensor field, i. e., that the BD scalar field is decoupled from the gravitational field. This is why GR is recovered in this limit. See, however, Ref. \cite{bd_faraoni_prd_1999}, where by means of the conformal transformations tool the author shows that the known result of Brans-Dicke theory reducing to general relativity when $\omega_\text{BD}\rightarrow\infty$, is false if the trace of the matter energy-momentum tensor vanishes.

For the general case with $V\neq 0$, following a procedure similar to the one exposed in Ref. \cite{stabile}, it is obtained the following expression for the effective (measured) gravitational coupling in the Brans-Dicke theory (see also \cite{salgado}):\footnote{This procedure is explained also in Ref. \cite{quiros-ijmpd-rev}.}

\bea 8\pi G_\text{eff}=\frac{1}{\phi_0}\left[\frac{3+2\omega_\text{BD}+e^{-M_0r}}{3+2\omega_\text{BD}}\right],\label{bd-v-eff-g}\eea where $\phi_0$ is the value of the field around which the perturbations are performed, while the mass (squared) of the propagating scalar perturbation is given by:

\bea M^2_0=\frac{\phi_0 V''_0}{3+2\omega_\text{BD}},\label{bd-mass-eff-g}\eea with $V_0=V(\phi_0)$, $V'_0=\der_\phi V|_{\phi_0}$, $V''_0=\der^2_\phi V|_{\phi_0}$, etc. It is seen that in the formal limit $M_0\rightarrow\infty$, i. e., when the propagating scalar degree of freedom decouples from the rest of the field spectrum of the theory, we recover general relativity with $8\pi G_\text{eff}=1$ (the choice $\phi_0=1$ is implicit). Meanwhile, in the limit of a light scalar field $M_0\rightarrow 0$, we retrieve the expression \eqref{bd-eff-g} for the measured gravitational coupling in the original formulation of the BD theory \cite{bd-1961}.


\subsection{Scalar-tensor theories}

In the case of the more general scalar-tensor theories given by the action principle:

\bea S=\int d^4x\sqrt{|g|}\left[\phi R-\frac{\omega(\phi)}{\phi}(\der\phi)^2-2V(\phi)+2{\cal L}_m\right],\label{stt-action}\eea for the measured gravitational coupling one gets:

\bea 8\pi G_\text{eff}=\frac{1}{\phi_0}\left[\frac{3+2\omega_0+e^{-M_0 r}}{3+2\omega_0}\right],\label{stt-gn}\eea where $\omega_0=\omega(\phi_0)$ and the mass $M_0$ of the scalar perturbation is given by \eqref{bd-mass-eff-g} with the replacement, $\omega_\text{BD}\rightarrow\omega_0$. The following formal limits (we consider the choice $\phi_0=1$): $M_0\rightarrow\infty$ and $M_0\rightarrow 0$, lead to general relativity and to massless STT, respectively.

For the particular case when the STT is given in the alternative form,\footnote{It is not difficult to prove that the action \eqref{stt-action} is transformed into \eqref{stt-alt-action} by a simple redefinition of the scalar field and of the coupling function: $$f(\phi)\rightarrow\phi,\;\;\omega(\phi)\rightarrow\frac{f(\phi)}{(\der_\phi f)^2}\,\omega(\phi).$$} 

\bea S=\int d^4x\sqrt{|g|}\left[f(\phi)R-\omega(\phi)(\der\phi)^2-2V(\phi)+2{\cal L}_m\right],\label{stt-alt-action}\eea the gravitational coupling that is measured in Cavendish-type experiments is given by:

\bea 8\pi G_\text{eff}=\frac{1}{f(\phi)}\left[\frac{4+2f/(\der_\phi f)^2}{3+2f/(\der_\phi f)^2}\right],\label{nmc-eff-g}\eea where, for simplicity, we have assumed vanishing self-interaction potential. 

We end up this section with a brief partial conclusion: In the expressions for the effective gravitational coupling in equations \eqref{bd-v-eff-g} and \eqref{stt-gn}, the factor $\phi^{-1}$ is associated with the tensor component of the gravitational interactions, while the factor $$\frac{3+2\omega+e^{-Mr}}{3+2\omega},$$ is originated from the scalar field contribution to the gravitational interactions. Hence, an adequate measure to determine whether a given theory is a STT or is just GR (with an additional scalar field matter source) may be, precisely, the effective gravitational coupling. While this may be a trivial exercise in very well known situations, for more complex cases where there are implied higher-order derivatives of the scalar field and/or derivative self-couplings, differentiating scalar-tensor theories from general relativity can be a more difficult task.


\section{Horndeski theories of gravity}\label{sect-horn-theor}

Horndeski theories \cite{horndeski_gal, nicolis_gal, deffayet_vikman_gal, deffayet_deser_gal, deffayet_prd_2011, kazuya_gal, fab_4_prl_2012, clifton-phys-rept-2012, deffayet-rev, tsujikawa_lect_not, kazuya_rpp_2016, quiros-ijmpd-rev, kobayashi-rpp-rev, odi-1, nojiri-rev, kase} represent further generalization of scalar-tensor theories to include higher derivatives of the scalar field and also derivative couplings. The history of the re-discovery of the Horndeski theories is quite peculiar. Inspired by the five-dimensional Dvali-Gabadadze-Porratti (DGP) model \cite{dgp_plb_2000, deffayet(dgp)_prd_2002, luty_porrati_rattazzi_jhep_2003, nicolis_rattazzi_jhep_2004, lue(dgp)_prd_2004, koyama_maartens_jcap_2006, roy-rev}, in Ref. \cite{nicolis_gal} the authors derived the five Lagrangians that lead to field equations invariant under the Galilean symmetry $\der_\mu\phi\rightarrow\der_\mu\phi+b_\mu$ in the Minkowski space-time. The scalar field that respects the Galilean symmetry was dubbed ``galileon''. Each of the five Lagrangians leads to second-order differential equations, keeping the theory free from unstable spin-2 ghosts, and from the corresponding instability of the resulting theory. If the analysis in Ref. \cite{nicolis_gal} is generalized to the curved spacetime, then these Lagrangians need to be promoted to their covariant forms. This was done in Ref. \cite{deffayet_vikman_gal} where the authors derived the covariant Lagrangians ${\cal L}_i$ ($i=1,..., 5$) that keep the field equations up to second-order. In Ref. \cite{kobayashi} it was shown that these Lagrangians are equivalent to the ones discovered by Horndeski \cite{horndeski_gal}.

According to Refs. \cite{deffayet_vikman_gal}, the most general scalar-tensor theories in four dimensions having second-order motion equations are described by the linear combinations of the following Lagrangians (${\cal L}_1=M^3\phi$, where the constant $M$ has the dimension of mass):

\bea &&{\cal L}_2=K,\;{\cal L}_3 =-G_3(\Box\phi),\;{\cal L}_4=G_4 R+G_{4,X}\left[(\Box\phi)^2-(\nabla_\mu\nabla_\nu\phi)^2\right],\nonumber\\
&&{\cal L}_5=G_5 G_{\mu\nu}\nabla^{\mu}\nabla^\nu\phi-\frac{1}{6}G_{5,X}\left[(\Box\phi)^3-3\Box\phi(\nabla_\mu\nabla_\nu\phi)^2+2(\nabla_\mu\nabla_\nu\phi)^3\right],\label{horn-lags}\eea where $K=K(\phi,X)$ and $G_i=G_i(\phi,X)$ ($i=3,4,5$), are functions of the scalar field $\phi$ and its kinetic energy density $X=-(\der\phi)^2/2$, while $G_{i,\phi}$ and $G_{i,X}$, represent the derivatives of the functions $G_i$ with respect to $\phi$ and $X$, respectively. In the Lagrangian ${\cal L}_5$ above, for compactness of writing, we have adopted the same definitions used in Ref. \cite{kobayashi}:
 
\bea (\nabla_\mu\nabla_\nu\phi)^2:=\nabla_\mu\nabla_\nu\phi\nabla^\mu\nabla^\nu\phi,\;(\nabla_\mu\nabla_\nu\phi)^3:=\nabla^\mu\nabla_\alpha\phi\nabla^\alpha\nabla_\beta\phi\nabla^\beta\nabla_\mu\phi.\label{def}\eea

The general action for the Horndeski theories:

\bea S_H=\int d^4x\sqrt{|g|}\left({\cal L}_2+{\cal L}_3+{\cal L}_4+{\cal L}_5+{\cal L}_m\right),\label{horn-action}\eea where the ${\cal L}_i$ are given by \eqref{horn-lags} and ${\cal L}_m$ stands for the Lagrangian of the matter degrees of freedom, comprises several well-known particular cases \cite{tsujikawa_lect_not}:

\begin{itemize}

\item{\it General relativity with a minimally coupled scalar field.} This is given by the following choice of the relevant functions in \eqref{horn-lags}:
 
\bea &&G_4=\frac{1}{2},\;G_3=G_5=0\nonumber\\
&&S=\int d^4x\sqrt{|g|}\left[\frac{1}{2}\,R+K(\phi,X)+{\cal L}_m\right].\label{k-ess-class}\eea This choice comprises quintessence; $K(\phi,X)=X-V$, and k-essence, for instance, $K(\phi,X)=f(\phi)g(X)$, where $f$ and $g$ are arbitrary functions of their arguments.

\item{\it Brans-Dicke theory.} The following choice corresponds to the BD theory \cite{bd-1961}:

\bea &&K(\phi,X)=\frac{\omega_\text{BD}}{\phi}\,X-V(\phi),\;G_3=G_5=0,\;G_4=\frac{\phi}{2},\nonumber\\
&&S=\frac{1}{2}\int d^4x\sqrt{|g|}\left[\phi R-\frac{\omega_\text{BD}}{\phi}(\der\phi)^2-2V\right].\nonumber\eea

\item{\it Non-minimal coupling (NMC) theory.} This is described by the functions:

\bea &&K=\omega(\phi)X-V(\phi),\;G_4=\frac{1-\xi\phi^2}{2},\;G_3=G_5=0,\nonumber\\
&&S=\int dx^4\sqrt{|g|}\left[\frac{1-\xi\phi^2}{2}R-\frac{\omega(\phi)}{2}(\der\phi)^2-V(\phi)\right].\nonumber\eea Higgs inflation \cite{bezrukov_plb_2008, bezrukov_jhep_2009} corresponds to the choice: $\omega(\phi)=1$, $V(\phi)=\lambda(\phi^2-v^2)^2/4$.

\item{\it Cubic galileon in the Jordan frame.} For this particular case in the functions in \eqref{horn-lags} one sets: $K=2\omega_\text{BD}X/\phi-2\Lambda\phi,$ $G_3=-2f(\phi)X,$ $G_4=\phi,\;G_5=0,$ and the resulting Jordan frame (JF) action reads \cite{kazuya_gal}: 

\bea S=\int d^4x\sqrt{|g|}\left[\phi R-\frac{\omega_\text{BD}}{\phi}(\der\phi)^2-2\Lambda\phi+f(\phi)\Box\phi(\der\phi)^2\right].\label{jf-qbic-gal}\eea

\item{\it Cubic galileon in the Einstein's frame.} For the choice: $$G_4=\frac{1}{2},\;G_5=0,\;G_3=2\sigma X,$$ where the constant $\sigma$ is the self-coupling parameter, we get the cubic galileon action in the Einstein's frame (EF) \cite{genly, dearcia}:

\bea S=\int\frac{d^4x\sqrt{|g|}}{2}\left\{R-\left[1+\sigma(\Box\phi)\right](\nabla\phi)^2-2V\right\}+\int d^4x\sqrt{-g}\,{\cal L}_m.\label{ef-qbic-gal}\eea The above action can be obtained from the Jordan frame (JF) action \eqref{jf-qbic-gal} through a disformal transformation \cite{disf-t-beken, disf-t-bruneton, disf-t-appleby, disf-t-bettoni, disf-t-kim, disf-t-rua, disf-t-arroja, disf-t-tsuji, disf-t-achour}.

\item{\it Kinetic coupling to the Einstein's tensor.} This is another particular and very interesting case within the class of the Horndeski theories \cite{sushkov, saridakis-sushkov, sushkov-a, k-coup-skugoreva, matsumoto, granda, gao, germani-prl, germani}. It corresponds to the following choice: $$K=X-V,\;G_3=0,\;G_4=\frac{1}{2},\;G_5=-\frac{\alpha}{2}\phi,$$ that leads to the action: 

\bea S=\frac{1}{2}\int d^4x\sqrt{|g|}\left[R+2(X-V)+\alpha G_{\mu\nu}\der^\mu\phi\der^\nu\phi\right],\label{k-coup-action}\eea where we have taken into account that integration by parts of the term $-\alpha\phi G_{\mu\nu}\nabla^\mu\nabla^\nu\phi$ amounts to $-\alpha G_{\mu\nu}\der^\mu\phi\der^\nu\phi$.
 
\end{itemize}


\subsection{Effective gravitational coupling in the Horndeski theories}\label{sect-horn-geff}

In order to compute the effective gravitational coupling let us consider perturbative expansion of the Horndeski motion equations around the flat Minkowski background space with metric $\eta_{\mu\nu}$, with constant value $\Phi$ of the scalar field \cite{hohmann_prd_2015}:

\bea g_{\mu\nu}=\eta_{\mu\nu}+h_{\mu\nu},\;\phi=\Phi+\psi,\;X=-\frac{1}{2}(\der\psi)^2.\label{perts-around-minkowski}\eea It is assumed, also, that the background is homogeneous, isotropic and stationary. Besides, since we will be interested in a single point mass source, the spherically symmetric solution will be considered. The computations of Ref. \cite{hohmann_prd_2015} are performed in the parametrized post-Newtonian (PPN) approximation. Under the former assumptions, the expression for the Newton's constant that is measured in Cavendish-like experiments is given by \cite{hohmann_prd_2015}:

\bea 8\pi G_\text{eff}=\frac{1}{2G_4}\left[\frac{3+2\omega_\text{eff}+e^{-M_\psi r}}{3+2\omega_\text{eff}}\right],\label{newton-c}\eea where

\bea M_\psi=\sqrt\frac{-2K_{,\phi\phi}}{K_{,X}-2G_{3,\phi}+3G_{4,\phi}^2/G_4},\label{m-psi}\eea is the mass of the scalar perturbation around the background value $\Phi$ and

\bea \omega_\text{eff}=\frac{G_4\left(K_{,X}-2G_{3,\phi}\right)}{2G_{4,\phi}^2},\label{weff-horn}\eea is the effective coupling of the scalar field to the curvature. In \eqref{newton-c}, \eqref{m-psi} and \eqref{weff-horn}, the coefficients $K$, $G_3$, $G_4$ and $G_5$ and their $\phi$ and $X$-derivatives are evaluated at background values: $\phi=\Phi$ and $X=0$.

The equation \eqref{newton-c} is the generalization of \eqref{stt-gn} for the case when higher-order derivatives of the scalar field are considered. Notice also that, in the formal limit when, $G_{4,\phi}\rightarrow 0\;\Rightarrow\;G_4=\text{const.},$ general relativity is recovered. In the form in \eqref{newton-c}, the above definition of the measured Newton's constant is not useful when $G_4=1/2$ is a constant since $\omega_\text{eff}$ is undefined. In this case we have to rewrite \eqref{newton-c} in the following equivalent way:

\bea 8\pi G_\text{eff}=\frac{1}{2G_4}\left[\frac{3G_{4,\phi}^2+G_4\left(K_{,X}-2G_{3,\phi}\right)+G_{4,\phi}^2e^{-M_\psi r}}{3G_{4,\phi}^2+G_4\left(K_{,X}-2G_{3,\phi}\right)}\right].\label{newton-c'}\eea From this equation it is seen that when the coefficient $G_4$ is a constant ($G_4=1/2$): $G_\text{eff}=(8\pi)^{-1}$, so that GR is recovered. 

We want to underline that, as the authors of Ref. \cite{hohmann_prd_2015} say, the above equations, in particular \eqref{newton-c} or \eqref{newton-c'}, are valid for those Horndeski theories in which screening mechanisms -- like the Vainshtein screening -- do not play a significant role so that the standard PPN formalism can be applied.


\subsubsection{The cosmological perturbations' approach}

Although the measured gravitational 'constant' in Horndeski theories, $G_\text{eff}$, can be found through the above explained procedure, there is an alternative way in which $G_\text{eff}$ can be derived without involving the PPN formalism, so that those Horndeski theories where the Vainshtein screening is an important ingredient, may be considered. It is based on the cosmological perturbations approach. The linear perturbations about the flat FRW metric: $$ds^2=-(1+2\psi)dt^2-2\der_i\chi dtdx^i+a^2(t)(1+2\Phi)\delta_{ij}dx^idx^j,$$ where $\psi$, $\chi$, and $\Phi$ are the scalar metric perturbations, in the theory given by the action \eqref{horn-action}, were studied in Ref. \cite{defelice(horn_perts)_plb_2011}. The spatial gauge where the $g_{ij}$ is diagonal is assumed. The scalar field as well as the matter fields, are also perturbed: $\phi(t)\rightarrow\phi(t)+\delta\phi(t,{\bf x})$, $\rho_m\rightarrow\rho_m+\delta\rho_m$. Following Ref. \cite{defelice(horn_perts)_plb_2011}, for compactness of writing, let us to introduce the following useful quantities:

\bea {\cal F}_T\equiv 2\left[G_4-X\left(\ddot\phi G_{5,X}+G_{5,\phi}\right)\right],\;{\cal G}_T\equiv 2\left[G_4-2XG_{4,X}-X\left(H\dot\phi G_{5,X}-G_{5,\phi}\right)\right],\label{horn-perts-usef-quant}\eea and also, the expansion:

\bea &&\Theta=-\dot\phi XG_{3,X}+2H\left(G_4-4XG_{4,X}-4X^2G_{4,XX}\right)+\dot\phi\left(G_{4,\phi}+2XG_{4,\phi X}\right)\nonumber\\
&&\;\;\;\;\;\;\;\;\;\;\;\;\;\;\;\;\;\;\;\;\;\;\;\;-H^2\dot\phi\left(5XG_{5,X}+2X^2G_{5,XX}\right)+2HX\left(3G_{5,\phi}+2XG_{5,\phi X}\right).\label{horn-expansion}\eea 

For the discussion on the evolution of matter perturbations relevant to large-scale structure, the modes deep inside the Hubble radius ($k^2/a^2\gg H^2$) are the ones that play the most important role. In the quasi-static approximation on sub-horizon scales\footnote{The range of validity of the quasi-static approximation may be very limited in theories where the sound speed $c_s\ll 1$.}, so that the dominant contributions in the perturbation equations are those including $k^2/a^2$ and $\delta$ -- the density contrast of matter, the following Poisson equation on $\psi$ is obtained \cite{defelice(horn_perts)_plb_2011}: $$\frac{k^2}{a^2}\,\psi\simeq-4\pi G_\text{eff}\delta\rho_m,$$ where the effective gravitational coupling $G_\text{eff}$, is the one measured in local experiments. It is given by the following expression (recall that we are working in the units system where $8\pi G_N=M_\text{pl}^{-2}=1$):

\bea 8\pi G_\text{eff}=\frac{2\left(B_6D_9-B_7^2\right)\left(\frac{k}{a}\right)^2-2B_6M^2}{\left(B_8^2D_9+A_6^2B_6-2A_6B_7B_8\right)\left(\frac{k}{a}\right)^2-B_8^2M^2},\label{horn-eff-g}\eea where 

\bea &&A_6=\frac{2(\Theta-H{\cal G}_T)}{\dot\phi},\;B_6=2{\cal F}_T,\;B_7=\frac{2\left[\dot{\cal G}_T+H\left({\cal G}_T-{\cal F}_T\right)\right]}{\dot\phi},\nonumber\\
&&B_8=2{\cal G}_T,\;D_9=\frac{2(\dot\Theta+H\Theta)-4H\dot{\cal G}_T+2H^2({\cal F}_T-2{\cal G}_T)+\rho_m}{\dot\phi^2}.\nonumber\eea The coefficient $M^2$ is related with the mass squared of the field $\delta\phi$ and it is given by:

\bea M^2=-K_{,\phi\phi}+K_{,\phi X}(\ddot\phi+3H\dot\phi)+2XK_{,\phi\phi X}+2XK_{,\phi XX}\ddot\phi+...,\label{M2}\eea where the ellipsis stands for terms containing second, third and fourth-order derivatives of the functions $G_i$ on the variables $\phi$ and $X$. For the full expression of $M^2$ see Eq. (35) of Ref. \cite{defelice(horn_perts)_plb_2011}.

In this paper Eq. \eqref{horn-eff-g} will be the master equation for determining the measured Newton's constant in Horndeski theories. Although the equation \eqref{newton-c} -- or in its equivalent form \eqref{newton-c'} -- serves the same purpose, as we have underlined above, these are based on the assumption that the PPN approximation is valid, so that \eqref{newton-c}, \eqref{newton-c'}, are not useful in those Horndeski theories where the Vainshtein screening (or other screening mechanisms) plays an important part. We shall further discuss on this issue in section \ref{sect-discu}.


\section{What are scalar-tensor theories?}\label{sect-horn-sst}

In the bibliography one usually finds the statement that the Horndeski theories are a generalization -- or an extension -- of the scalar-tensor theories. But, what really means that a given theory of gravity is a scalar-tensor theory? Here such a statement entails that the gravitational phenomena are not completely due to the curvature of spacetime but, that these are partly a result of the curvature and partly due to an additional scalar field degree of freedom. In other words, the perturbative spectrum of the theory consists of 3 d.o.f.: the two polarizations of the graviton and the scalar perturbation. These are the degrees of freedom that carry the gravitational interactions in contrast to the two degrees of freedom that propagate the gravitational effects in GR. Take as an example a scalar field with the typical non-minimal coupling to the curvature of the form, ${\cal L}_\text{nmc}\propto f(\phi)R$. In this case the gravitational coupling $\propto f^{-1}(\phi)$, so that it sets the strength of the gravitational interactions at each point in spacetime. This is the most obvious way in which the scalar field modifies the gravitational interactions. In addition, the measured (effective) gravitational coupling is modified in a non-trivial way. For instance, if the scalar field possesses a standard kinetic term, $-(\der\phi)^2/2$, the above non-minimal coupling implies that the measured gravitational constant is given by \eqref{nmc-eff-g}. For vanishing kinetic term without the potential the scalar field is a non-propagating degree of freedom, so that the resulting theory coincides with general relativity. But if the scalar field's potential is non-vanishing, it could happen that for vanishing kinetic term the theory is a scalar-tensor one.

As mentioned in the concluding paragraph of section \ref{sect-bd}, a good indicator that the given theory is a STT is that its corresponding effective gravitational coupling be a function of the scalar field, i. e., that it could be expressible in the form of \eqref{stt-gn} through, possibly, a redefinition of the scalar field. After the Horndeski generalizations of the scalar-tensor theories, one should require that, not only the scalar field but also its higher order derivatives and mixed (non-linear) terms where curvature quantities are multiplied by these elements, can modify the effective coupling that is measured in Cavendish-like experiments \eqref{horn-eff-g}. For the Brans-Dicke theory, for instance: $$K(\phi,X)=\frac{\omega_\text{BD}}{\phi}\,X-V(\phi),\;G_3=G_5=0,\;G_4=\frac{\phi}{2}.$$ The corresponding effective gravitational coupling \eqref{horn-eff-g} is given by:

\bea 8\pi G_\text{eff}=\frac{1}{\phi}\left[\frac{4+2\omega_\text{BD}+2\phi(Ma/k)^2}{3+2\omega_\text{BD}+2\phi(Ma/k)^2}\right],\label{horn-bd-eff-g}\eea where, neglecting terms ${\cal O}(H^2\phi)$, $M^2\simeq\der^2_\phi V+\der_\phi V/\phi$. In the limit $M^2\rightarrow 0$, i. e., when the scalar field is massless as in the original BD theory without the potential, we recover the known result of \eqref{bd-eff-g}. Meanwhile, in the limit $M^2\rightarrow\infty$, i. e., when the scalar field decouples from the rest of the matter degrees of freedom of the theory -- also when $\omega_\text{BD}\rightarrow\infty$ -- the GR behavior is reproduced.

But, what about other theories included in the Horndeski class? Take, for instance, the class determined by the choice \eqref{k-ess-class}. Looking at the resulting action, for an arbitrary function $K(\phi,X)$, one immediately recognizes the so called k-essence theories (these include the quintessence models for the particular choice $K(\phi,X)=X-V(\phi)$). In this case, since $G_4=1/2$, ${\cal F}_T={\cal G}_T=1$, and given that $G_3=G_5=0$, one gets that $\Theta=H$, and consequently, $A_6=B_7=0$, $B_6=B_8=2$. Hence, for the effective gravitational coupling \eqref{horn-eff-g} one obtains $8\pi G_\text{eff}=1$, which means that k-essence is just general relativity plus a scalar field -- with a perhaps exotic kinetic energy term --  as matter source of the Einstein's equations. It is not a scalar-tensor theory.


\subsection{The 'Einstein frame' cubic galileon}\label{subsect-qbic-gal}

For the choice:

\bea &&G_4=\frac{1}{2},\;G_5=0,\;G_3=G_3(\phi,X)\neq 0,\nonumber\\
&&S=\int d^4x\sqrt{|g|}\left[\frac{1}{2}\,R-G_3(\phi,X)\Box\phi\right],\label{qbic-class}\eea that includes the EF cubic galileon model \eqref{ef-qbic-gal}, we have that ${\cal F}_T={\cal G}_T=1$, while $\Theta=H-\dot\phi XG_{3,X}$, and 

\bea A_6=2(\Theta-H)/\dot\phi,\;B_6=B_8=2,\;D_9=[2\dot\Theta+2H(\Theta-H)+\rho_m]/\dot\phi^2,\nonumber\eea so that

\bea 8\pi G_\text{eff}=\frac{[2\dot\Theta-2H\dot\phi XG_{3,X}+\rho_m]\left(\frac{k}{a}\right)^2-M^2\dot\phi^2}{[2\dot\Theta-2H\dot\phi XG_{3,X}+4X^3G^2_{3,X}+\rho_m]\left(\frac{k}{a}\right)^2-M^2\dot\phi^2}.\label{qbic-class-eff-g}\eea Notice that if, $G_3=G_3(\phi)$, is a function of the scalar field only, the resulting theory is equivalent to GR.\footnote{As a matter of fact, a term $\propto G_3(\phi)\Box\phi$ in the Lagrangian density may be integrated by parts to give $2G_{3,\phi}\,X$, which may be absorbed in the $K(\phi,X)$-term, so that the resulting theory is given by \eqref{k-ess-class}.} In order for the above choice to represent a STT it is required that $G_3$ be an explicit function of the kinetic term $X$.

The cubic galileon represents an example where the scalar-tensor character of a given theory may be very subtle. Actually, for the choice \eqref{k-ess-class} it is clear why the resulting theory is general relativity with a scalar field as matter source: there is no direct coupling of the scalar field (or of its derivatives) to the curvature. These couplings are explicit in the terms: $$G_4(\phi,X)R,\;\;G_5(\phi,X)G_{\mu\nu}\nabla^\mu\nabla^\nu\phi,$$ but as long as $G_4=\text{const}=1/2$ and $G_5=0$, there is no (explicit) direct coupling between the scalar field a the curvature. The interesting thing is that according to the choice \eqref{qbic-class}, $G_4=1/2$, $G_5=0$, as in \eqref{k-ess-class}, so that one should expect that the resulting theory should be general relativity as well. However, if take a closer look at \eqref{qbic-class-eff-g}, it is seen that thanks to the term $4X^3G^2_{3,X}$ in the denominator, the Newton's constant measured in Cavendish-type experiments is a function of the spacetime point through the field variables and their derivatives: $$G_\text{eff}\propto f(H,\dot H,\phi,X,\dot X),$$ so that this is not general relativity but a scalar-tensor theory! 

We may explain the above result in the following way. For simplicity let us assume that $G_3=G_3(X)$ is an explicit function of the kinetic energy of the scalar field only. Variation of the Lagrangian ${\cal L}_3$ in \eqref{horn-lags} with respect to the scalar field can be written as: 

\bea \delta{\cal L}_3=-G_{3,X}\delta X(\Box\phi)-G_3\Box(\delta\phi),\nonumber\eea where $\delta X=\nabla^\mu\phi\nabla_\mu(\delta\phi)$. Up to a divergence the variation of the Lagrangian can be put into the following form:

\bea \delta{\cal L}_3=\left[G_{3,XX}\nabla_\mu X\nabla^\mu\phi(\Box\phi)+G_{3,X}(\Box\phi)^2+G_{3,X}\nabla^\mu\phi\nabla_\mu(\Box\phi)-G_{3,XX}\nabla_\mu X\nabla^\mu X-G_{3,X}\Box X\right]\delta\phi,\nonumber\eea where the presence of third-order derivatives is evident. By means of the relationship: 

\bea \Box(\nabla_\mu\phi)-\nabla_\mu(\Box\phi)=R_{\mu\nu}\nabla^\mu\phi\nabla^\nu\phi,\label{curv-der-rel}\eea the variation of the cubic Lagrangian can be finally rewritten into the form where it contains derivatives no higher than the 2nd order \cite{quiros-ijmpd-rev}:

\bea \delta{\cal L}_3=\left\{G_{3,XX}\nabla_\mu X\left[\nabla^\mu\phi(\Box\phi)-\nabla^\mu X\right]+G_{3,X}\left[(\Box\phi)^2-(\nabla_\mu\phi\nabla_\nu\phi)^2\right]-G_{3,X}R_{\mu\nu}\nabla_\mu\phi\nabla_\nu\phi\right\}\delta\phi.\label{delta-lag}\eea This has been achieved at the cost of introducing a term (last term above) where the Ricci curvature tensor is coupled to the derivatives of the scalar field. This makes evident that any first-order variation of the cubic galileon Lagrangian induces a derivative coupling of the scalar field to the curvature, thus making explicit the scalar-tensor character of the cubic galileon theory.


\subsection{Theory with kinetic coupling to the Einstein's tensor}

As stated in section \ref{sect-horn-theor}, the choice, $$K=X-V,\;G_3=0,\;G_4=\frac{1}{2},\;G_5=-\frac{\alpha}{2}\phi,$$ results in the gravitational theory with kinectic coupling to the Einstein's tensor that is given by the action \eqref{k-coup-action}. In this case \eqref{horn-eff-g} is written as:

\bea 8\pi G_\text{eff}=\frac{\left\{{\cal F}_T{\cal H}_T+{\cal J}_T^2\right\}\left(\frac{k}{a}\right)^2+{\cal F}_TM^2X}{\left\{{\cal G}_T^2{\cal H}_T-{\cal F}_T\left({\cal F}_T-{\cal G}_T\right)^2H^2-2{\cal G}_T\left({\cal F}_T-{\cal G}_T\right){\cal J}_TH\right\}\left(\frac{k}{a}\right)^2+{\cal G}_T^2M^2X},\label{k-coup-gn}\eea where

\bea &&{\cal F}_T=2G_4+\alpha X,\;{\cal G}_T=2G_4-\alpha X,\nonumber\\
&&{\cal H}_T\equiv\left({\cal F}_T-2{\cal G}_T\right)\dot H+{\cal F}_TH,\;{\cal J}_T\equiv\dot{\cal G}_T+H\left({\cal F}_T-{\cal G}_T\right).\label{k-coup-misc}\eea

That this choice is a scalar-tensor theory -- as corroborated by \eqref{k-coup-gn} where it is apparent that $G_\text{eff}=f(H,\dot H,X,\dot X)$ is a function of the spacetime point -- is evident from the action \eqref{k-coup-action} where the coupling of the derivatives of the scalar field to the Einstein's tensor, $G_{\mu\nu}\der^\mu\phi\der^\nu\phi$, is explicit.


\section{$f(R)$ theories as scalar-tensor theories}\label{sect-fdr}

It is well-known that under certain conditions the $f(R)$ theories are mathematically equivalent to the BD theory \cite{nojiri-rev, fdr-sotiriou-cqg, fdr-sotiriou-faraoni-rev, fdr-tsujikawa-lrr}. This is true no matter which formalism to follow: either the metric or the Palatini formalism. According to the metric formalism given by the action:

\bea S_\text{met}=\frac{1}{2}\int d^4x\sqrt{|g|}\left[f(R)+{\cal L}_m\left(g_{\mu\nu},\psi\right)\right],\label{met-fdr-action}\eea where $f(R)$ is a function of the Ricci scalar $R$ and ${\cal L}_m\left(g_{\mu\nu},\psi\right)$ stands for the Lagrangian of the matter degrees of freedom $\psi$, the given $f(R)$ theory is equivalent to JFBD theory with vanishing coupling parameter $\omega_\text{BD}=0$ and with the potential $V$. Actually if introduce the auxiliary field $\chi$ the gravitational part of the above action can be written in the following way

\bea S_\text{met}=\frac{1}{2}\int d^4x\sqrt{|g|}\left[f'(\chi)\left(R-\chi\right)+f(\chi)\right].\label{aux}\eea By varying this action with respect to the auxiliary field one obtains that: $f''(\chi)(R-\chi)=0$, so that if $f''(\chi)\neq 0$, then $\chi=R$ and the action \eqref{met-fdr-action} is retrieved. If redefine $\phi=f'(\chi)$ and introduce the following definition: $V(\phi)=\chi(\phi)\,\phi-f(\chi(\phi))$, the gravitational part of the action \eqref{met-fdr-action} can be written in the following form:

\bea S_\text{met}=\frac{1}{2}\int d^4x\sqrt{|g|}\left[\phi R-V(\phi)+{\cal L}_m\left(g_{\mu\nu},\psi\right)\right].\label{met-bd-action}\eea Hence, mathematical equivalence of the actions \eqref{met-fdr-action} and \eqref{met-bd-action} requires that $f''(\chi)\neq 0$, which implies that $\phi=f'(R)$.

Within the framework of the Palatini formalism:

\bea S_\text{Pal}=\frac{1}{2}\int d^4x\sqrt{|g|}\left[f(\hat R)+{\cal L}_m\left(g_{\mu\nu},\psi\right)\right],\label{pal-fdr-action}\eea where geometric objects and operators, including the curvature scalar $\hat R$, are defined in terms of the affine connection: $$\Gamma^\lambda_{\mu\nu}=\{^\lambda_{\mu\nu}\}+\frac{1}{f'(\hat R)}\left[\delta^\lambda_\mu\der_\nu\left(f'(\hat R)\right)+\delta^\lambda_\nu\der_\mu\left(f'(\hat R)\right)-g_{\mu\nu}\der^\lambda\left(f'(\hat R)\right)\right],$$ with $\{^\lambda_{\mu\nu}\}$ -- the Christoffel symbols of the metric. We have that: 

\bea R=\hat R-\frac{3}{2\left[f'(\hat R)\right]^2}\left(\der f'(\hat R)\right)^2-\frac{3}{f'(\hat R)}\Box f'(\hat R).\label{r-rhat-rel}\eea If, following the same procedure than in the metric formalism, introduce an auxiliary field $\chi$ and make the appropriate identifications one obtains an action similar to \eqref{met-bd-action} with the substitution of $R$ by $\hat R$. Then, by considering \eqref{r-rhat-rel}, one obtains that \eqref{pal-fdr-action} is mathematically equivalent to the JFBD action with the singular value of the coupling parameter $\omega_\text{BD}=-3/2$:

\bea S_\text{Pal}=\frac{1}{2}\int d^4x\sqrt{|g|}\left[\phi R+\frac{3}{2\phi}(\der\phi)^2-V(\phi)+{\cal L}_m\left(g_{\mu\nu},\psi\right)\right].\label{pal-bd-action}\eea

In general (arbitrary coupling constant $\omega_\text{BD}$ and minimally coupled matter degrees of freedom $\psi$) the BD equations of motion read:

\bea &&G_{\mu\nu}=\frac{1}{\phi}\,T^{(m)}_{\mu\nu}-\frac{1}{2\phi}\,g_{\mu\nu}V(\phi)+\frac{\omega_\text{BD}}{\phi^2}\left[\der_\mu\phi\der_\nu\phi-\frac{1}{2}\,g_{\mu\nu}(\der\phi)^2\right]+\frac{1}{\phi}\left[\nabla_\mu\nabla_\nu-g_{\mu\nu}\Box\right]\phi,\nonumber\\
&&(3+2\omega_\text{BD})\Box\phi=T^{(m)}-2V(\phi)+\phi V_{,\phi},\label{bd-moteq}\eea where $T^{(m)}=g^{\mu\nu}T_{\mu\nu}^{(m)}$ is the trace of the stress-energy tensor of the matter degrees of freedom. As it is seen from these equations, the scalar field in \eqref{met-bd-action} is dynamical since for $\omega_\text{BD}=0$ there is a motion equation for $\phi$: $$3\Box\phi=T^{(m)}-2V(\phi)+\phi V_{,\phi}.$$ Meanwhile, the scalar field in \eqref{pal-bd-action} is not dynamical, since for $\omega_\text{BD}=-3/2$, the BD Klein-Gordon equation in \eqref{bd-moteq} amounts to an algebraic constraint and not to an equation of motion: 

\bea T^{(m)}=2V(\phi)-\phi V_{,\phi}.\label{const-eq}\eea By solving \eqref{const-eq} one may find $\phi=\phi(T^{(m)})$, but for the particular potential $V=\mu^2\phi^2$, in which case the latter equation amounts to $T^{(m)}=0$. For the quartic potential $V=\lambda\phi^4$, for instance, one gets that: $$\phi_\pm=\pm\left(-\frac{T^{(m)}}{2\lambda}\right)^{1/4}.$$ If substitute back the corresponding $\phi=\phi(T^{(m)})$ into the Einstein's equation in \eqref{bd-moteq} -- recall that in this case $\omega_\text{BD}=-3/2$ -- what one gets is a theory of gravity where the effect of matter on the geometry is enhanced as compared with general relativity.

If follow the weak-field approach in the metric formalism \eqref{met-fdr-action}/\eqref{met-bd-action}, it can be shown that the effective Newton's constant in this case reads \cite{olmo-prd-2005}:

\bea 8\pi G_N=\frac{1}{\phi_0}\left[1+\frac{1}{3}\,e^{-m_\phi r}\right],\label{met-eff-g}\eea where $r=|\bf{x}-\bf{x}'|$, the scalar field is evaluated today, $\phi_0=\phi(t_0)$ (it is determined by appropriate cosmological boundary conditions), $V_0=V(\phi_0)$ and 

\bea m^2_\phi=\frac{\phi_0 V''_0-V'_0}{3}\geq 0.\label{mass2}\eea If compare equation \eqref{met-eff-g} with \eqref{bd-v-eff-g} (with the substitution $\omega_\text{BD}=0$) it is seen that these coincide but for the definition of the mass parameters in \eqref{bd-mass-eff-g} and in \eqref{mass2}. In the definition \eqref{bd-mass-eff-g} the term with the first derivative $V'_0$, does not appear, because in \cite{salgado} and related bibliographic references, the value $\phi_0$ is associated with vacuum background value where $V'(\phi_0)=0$. 

For the Palatini approach the weak field limit has its own features given that the scalar field, as mentioned, is not dynamical. In this case the terms with $\phi=\phi(T^{(m)})$ should be treated as matter and expanded in post-Newtonian orders and not around the vacuum. In this case, for the measured Newton's constant one gets \cite{olmo-prd-2005}:

\bea 8\pi G_N=\frac{1}{\phi_0}\left(1+\frac{M_V}{M}\right),\;M=\int d^3x'\frac{\phi_0}{\phi}\,\rho(t,{\bf x}'),\;M_V=\int d^3x'\left(\frac{V_0}{\phi_0}-\frac{V}{\phi}\right)\phi_0,\label{pal-eff-g}\eea where $\phi_0$ is a solution of \eqref{const-eq} with $T^{(m)}=0$ and $V_0=f(R_0)$ with $R_0=\hat R\left(T^{(m)}=0\right)$. Notice that this equation can not be obtained from \eqref{bd-v-eff-g} since in the limit $\omega_\text{BD}\rightarrow-3/2$ the effective mass \eqref{bd-mass-eff-g} tends to infinitely large values, meaning that the scalar field is decoupled from the dynamics of gravity. I. e., from \eqref{bd-v-eff-g} it follows that in the $\omega_\text{BD}\rightarrow-3/2$ limit general relativity is recovered. Quite a different result is obtained if follow the procedure proposed in \cite{olmo-prd-2005} in connection with the Palatini formalism applied to the $f(R)$ action. In this case, according to \eqref{pal-eff-g}, the resulting theory is a modification of GR where the matter contribution to the curvature of spacetime is enhanced by the trace of the stress-energy tensor.


\section{Conformal transformations and classification of scalar-tensor theories}\label{sect-ct}

There is another aspect of the classification of the scalar-tensor theories of gravity that is related with the so called conformal transformations' issue. The issue can be stated in the following way: Under conformal transformations of the metric \eqref{conf-t} the given STT may be formulated in a -- in principle infinite -- set of mathematically equivalent field variables, called as conformal frames. Among these the Jordan frame (JF) and the Einstein's frame (EF) are the most outstanding \cite{dicke-1962, faraoni, faraoni_prd_2007, sarkar_mpla_2007, sotiriou, deruelle, quiros_grg_2013}. The following related questions are the core of the conformal transformation's issue.\footnote{The equivalence of different representations of scalar-tensor theories of gravity under the method of Legendre/conformal transformations including the boundary terms has been investigated in Ref. \cite{saltas}. The quantum equivalence of different representations was also briefly discussed in that reference.}

\begin{enumerate}

\item Are the different conformal frames not only mathematically equivalent but, also, physically equivalent?

\item If the answer to the former question were negative, then: which one of the conformal frames is the physical one, i. e., the one in terms of whose field variables to interpret the physical consequences of the theory? 

\end{enumerate} The resulting controversy originates from the lack of consensus among different researchers and also among the different points of view of the same researcher along her/his research history, regarding their answer to the above questions. There are known classifications of the different works of different authors and of the same author on this issue \cite{faraoni}. That the controversy has not been resolved yet is clear from the amount of yearly work on the issue where there is no agreement about the correct answer to these questions \cite{saal_cqg_2016, indios_consrvd_prd_2018, dimakis, thermod_prd_2018, hammad_prd_2019, ct-1, ct-2, ct(fresh-view)-3, christian, paliatha-2, ct(quant-equiv)-4, ct(quant-equiv)-5, ct(quant-equiv)-6, ct(inequiv)-7, ct(inequiv)-8, ct(inequiv)-9}. Here we shall approach to the conformal frames' issue from the classical standpoint exclusively. For a related discussion based on quantum arguments we recommend Refs. \cite{ct(quant-equiv)-4, ct(quant-equiv)-5, ct(quant-equiv)-6, ct(inequiv)-7, ct(inequiv)-8} and references therein. 


What does the classification of the STT-s have to say about the conformal transformations' issue? Under a conformal transformation of the metric \eqref{conf-t} with $\Omega^2=\phi$, and $\vphi=\ln\phi$, the JFBD action, 

\bea S_\text{JF}=\int d^4x\sqrt{|g|}\left[\phi R-\frac{\omega_\text{BD}}{\phi}(\der\phi)^2-2V(\phi)+2{\cal L}_m(\psi^{(i)},\nabla\psi^{(i)},g_{\mu\nu})\right],\label{jfbd-action}\eea is transformed into the EFBD theory with action:\footnote{Historically, after the well-known paper \cite{dicke-1962} by Dicke, it is usual to regard the JF as that formulation in which the scalar field is non-minimally coupled to the curvature scalar while it is minimally coupled to the matter d.o.f, meanwhile the EF is associated with the conformal formulation in which the scalar field is minimally coupled to the curvature scalar but non-minimally coupled to the matter sector. The requirement of minimal coupling of the scalar field to the matter Lagrangian in the JF was based on the expectations of the authors of \cite{bd-1961} that their theory obeyed both the Mach principle and the strong bonds on the violation of the weak equivalence principle. We want to notice, however, that other possibilities are allowed as well. For instance, the scalar field may be non-minimally coupled both to the Ricci scalar and to the matter d.o.f (see, for instance \cite{quiros-prd-2000}). Take, as an illustration, the theory where the scalar field is non-minimally coupled both to the curvature and to the matter Lagrangian in the following way (for simplicity we omit the self-interaction potential): $$S=\int d^4x\sqrt{|g|}\left[\phi R-\frac{\omega_\text{BD}}{\phi}(\der\phi)^2+2\phi^2{\cal L}_m(\psi^{(i)},\nabla\psi^{(i)},g_{\mu\nu})\right].$$ It can be shown that under the conformal transformation \eqref{conf-t} with $\Omega^2=\phi$, and $\vphi=\ln\phi$, the above action is mapped into standard GR with two minimally coupled matter fields: the scalar field and the standard matter d.o.f; $$S=\int d^4x\sqrt{|g|}\left[R-\left(\omega_\text{BD}+\frac{3}{2}\right)(\der\vphi)^2+2{\cal L}_m(\psi^{(i)},\nabla\psi^{(i)},e^{-\vphi}g_{\mu\nu})\right].$$ Notice, however, that the standard matter d.o.f is coupled to the conformal metric as in the EFBD. There are other scalar-tensor modifications of GR, such as those inspired in the string theory, where the different matter d.o.f couple to differnt conformal metrics, etc. The main conclusion is that not every possible scalar-tensor modification of GR can be brought in the JF or in the EF, at least not in the same way as the BD theory.} 

\bea S_\text{EF}=\int d^4x\sqrt{|g|}\left[R-\left(\omega_\text{BD}+\frac{3}{2}\right)(\der\vphi)^2-2V(\vphi)+2e^{-2\vphi}{\cal L}_m(\psi^{(i)},\nabla\psi^{(i)},e^{-\vphi}g_{\mu\nu})\right],\label{efbd-action}\eea where $e^{2\vphi}V(\vphi)=V(\phi)$. It is seen from \eqref{newton-c'} that for the JFBD theory the measured Newton's constant, 

\bea 8\pi G_\text{eff}=\frac{1}{\phi}\left[\frac{2\omega_\text{BD}+3+e^{-M_\psi r}}{2\omega_\text{BD}+3}\right],\;M_\psi=\frac{2}{\phi}\sqrt\frac{\phi^3V_{,\phi\phi}-2\omega_\text{BD}X}{2\omega_\text{BD}+3},\label{jfbd-geff}\eea is a function of the spacetime point, while in the framework of EFBD, $8\pi G_\text{eff}=1$, i. e., it is general relativity with an universal fifth-force: $$f_\mu^\text{5th}\propto-\frac{1}{2}\nabla_\mu\vphi,$$ that is originated from the non-minimal coupling of the dilaton field $\vphi$ with the matter Lagrangian in $S_\text{EF}$: $e^{-2\vphi}{\cal L}_m$. The way this fifth-force arises can be seen if realize that the time-like geodesics of the metric in the JF:\footnote{The null geodesics, i. e., the photons' paths, are not transformed by the conformal transformations.}

\bea \frac{d^2x^\mu}{ds^2}+\left\{^{\;\mu}_{\sigma\lambda}\right\}\frac{dx^\sigma}{ds}\frac{dx^\lambda}{ds}=0,\nonumber\eea are transformed by the conformal transformations into non-geodesics in the EF:

\bea \frac{d^2x^\mu}{d\tau^2}+\left\{^{\;\mu}_{\sigma\lambda}\right\}\frac{dx^\sigma}{d\tau}\frac{dx^\lambda}{d\tau}=-\frac{1}{2}\nabla^\mu\vphi,\nonumber\eea where the affine reparametrization $ds\rightarrow\Omega^{-1}d\tau$ has been performed. In the presence of matter fluxes the above transformation property is reflected in the conformal transformation of the continuity equation:

\bea \nabla^\mu T^{(m)}_{\mu\nu}=0\;\rightarrow\;\nabla^\mu T^{(m)}_{\mu\nu}=-\frac{1}{2}T^{(m)}\nabla_\mu\vphi.\nonumber\eea

The two conformally related formulations of the BD theory represent, to all purposes, two different theories: i) JFBD is a STT while EFBD is GR with an additional non-gravitational (universal) fifth-force acting on the matter fields, and ii) while the weak equivalence principle (WEP) is valid in the JFBD metric theory of gravity, it is violated in the EFBD non-metric theory. According to our own point of view, physical equivalence requires of an underlying symmetry \cite{quiros_grg_2013, sc_inv_jackiw, quiros-arxiv}. For instance, invariance under general coordinate transformations and/or under gauge transformations, warrants physical equivalence of the different coordinate systems and/or of the different gauges. The quantities having the physical meaning are those which are not transformed by the coordinate and/or gauge transformations. In a similar way, the conformally related (thus mathematically equivalent) frames are physically equivalent only if the given STT -- and the corresponding field equations -- is invariant under the conformal transformation (plus an innocuous rescaling of the scalar field). In this case conformal invariance is the symmetry that underlies the physical equivalence of the different representations of the theory.\footnote{An example of such a truly conformal invariant theory is explored in the second reference in \cite{quiros-arxiv}.} The quantities that have the physical meaning are those that are invariant under the conformal transformation (plus general coordinate transformations). As an example of a theory that is invariant under the conformal transformations so that the different conformal frames are actually physically equivalent, lets to choose a conformally coupled scalar field like in the following action:\footnote{Under the following redefinition of the scalar field, $\phi^2\rightarrow 6\exp\vphi$, this action can be written in terms of the so called ``string frame'' variables: $$S=\frac{1}{2}\int d^4x\sqrt{|g|}\,e^\vphi\left[R+\frac{3}{2}(\der\vphi)^2\pm 3\lambda\,e^\vphi\right].$$ The problem with this theory which, as seen is BD with the anomalous value of the coupling constant $\omega_\text{BD}=-3/2$, is that only massless fields can be coupled to gravity in a consistent way.}

\bea S=\int d^4x\sqrt{|g|}\left[\frac{\phi^2}{12}\,R+\frac{1}{2}(\der\phi)^2\pm\frac{\lambda}{12}\,\phi^4\right].\label{deser-action}\eea Under the Weyl rescalings:

\bea g_{\mu\nu}\rightarrow\Omega^{-2}g_{\mu\nu},\;\phi\rightarrow\Omega\,\phi,\label{scale-t}\eea the combination $\sqrt{|g|}[\phi^2R+6(\der\phi)^2]$, as well as the scalar density, $\sqrt{|g|}\,\phi^4$, are kept unchanged. Then, if further assume that the dimensionless constant $\lambda$ is not transformed by the Weyl rescalings, the action (\ref{deser-action}) is invariant under \eqref{scale-t}. Any scalar field which appears in the gravitational action the way $\phi$ does in \eqref{deser-action}, is said to be conformally coupled to gravity. Hence, for instance, the following action \cite{sc_inv_jackiw, quiros-arxiv, sc_inv_prester, sc_inv_padilla, sc_inv_bars, sc_inv_bars_1, sc_inv_bars_2, sc_inv_carrasco, sc_inv_alpha, sc_inv_alpha_2, sc_inv_alpha_3, sc_inv_alpha_4}:

\bea S=\int d^4x\sqrt{|g|}\left[\frac{\left(\phi^2-\sigma^2\right)}{12}\,R+\frac{1}{2}(\der\phi)^2-\frac{1}{2}(\der\sigma)^2\right],\label{bars-action}\eea is also invariant under \eqref{scale-t} since both $\phi$ and $\sigma$ are conformally coupled to gravity,\footnote{For the coupling $\propto (\phi^2-\sigma^2)^{-1}$ to be positive and the theory Weyl-invariant, the scalar $\phi$ must have a wrong sign kinetic energy -- just like in \eqref{deser-action} -- potentially making it a ghost. However, the local Weyl gauge symmetry compensates, thus ensuring the theory is unitary \cite{sc_inv_bars, sc_inv_bars_1, sc_inv_bars_2}.} provided that the additional scalar field $\sigma$ transforms in the same way as $\phi$: $\sigma\rightarrow\Omega\,\sigma$. Hence, the above theories \eqref{deser-action}, \eqref{bars-action}, embody the discussed physical equivalence of the conformal frames. 

The above examples show a different scene than the one for the STT-s, including the BD theory. Under a conformal transformation the latter theories are mapped from one frame (say the JF) into a different frame (say the EF), with different equations of motion and, consequently, different physical implications. This means that the JF and EF (and other conformal frames) in which a given STT may be formulated are not physically equivalent despite of their mathematical equivalence. 

Nevertheless, one of the most widespread points of view on the conformal transformations' issue found in the bibliography (see, for instance, Refs. \cite{dicke-1962, faraoni, faraoni_prd_2007, sotiriou}), assumes that the JFBD and the EFBD theories are physically equivalent representations of given gravitational phenomena. If this point of view were correct, then, metric STT and non-metric GR theories of gravity were indistinguishable. I. e. there were no point in classifying theories in STT and GR or in metric an non-metric, bimetric, etc. In this paper we want to be clear in that, although we do not share the point of view stating equivalence of the conformal frames, the issue is still under debate \cite{ct-equiv, ct-inequiv}. Only if one does not believe in the conformal equivalence, an unambiguous classification of gravitational theories into STT-s and GR is possible.


\subsection{$f(R)$ theories}

Given the mathematical equivalence between $f(R)$ theories and JFBD theory -- see section \ref{sect-fdr} -- the question about the physical equivalence of the different conformal frames (in particular of the JF and of the EF) in which the BD theory may be formulated, also arises in the framework of $f(R)$ theories \cite{fdr-sotiriou-cqg}. In this case, as shown in \ref{sect-fdr}, the Jordan frame $f(R)$ theory \eqref{met-fdr-action} is mathematically equivalent to JFBD theory \eqref{met-bd-action} if $f''(\chi)\neq 0$. Under a conformal transformation of the metric, $g_{\mu\nu}\rightarrow\left(f'(\chi)\right)^{-1}g_{\mu\nu}$, together with the redefinition, $\vphi=-\ln\left[f'(\chi)\right]$, the JFBD theory \eqref{met-fdr-action}/\eqref{met-bd-action} is transformed into the Einstein's frame \cite{nojiri-rev, rev-2-1, rev-2-2}:

\bea S_\text{EF}=\frac{1}{2}\int d^4x\sqrt{|g|}\left[R-\frac{3}{2}(\der\vphi)^2-V(\vphi)\right],\label{2-ef-action}\eea where $$V(\vphi)=\frac{\chi}{f'(\chi)}-\frac{f(\chi)}{\left[f'(\chi)\right]^2}.$$ Conversely, GR with a scalar field as matter source: $$S_\text{GR}=\frac{1}{2}\int d^4x\sqrt{|g|}\left[R-(\der\phi)^2-2V(\phi)\right],$$ can be written in the form of a $f(R)$ theory \cite{nojiri-rev}. This true only for scalar fields with a canonical kinetic term. For ghost scalar fields this inverse mapping is not possible.

In the EF formulation \eqref{2-ef-action} the matter Lagrangian ${\cal L}_m$ acquires a non-minimal coupling of the following form: $${\cal L}_m\left(g_{\mu\nu},\psi\right)\rightarrow\left(f'(\chi)\right)^{-2}{\cal L}_m\left(\left(f'(\chi)\right)^{-1}g_{\mu\nu},\psi\right).$$ This means that the matter fields do not follow geodesics of the metric since these feel an additional non-gravitational interaction that deviates their motion from being geodesic. This, as well as other results in connection to the discussion on the physical equivalence between the JF and EF formulations of $f(R)$ gravity theories as, for instance, the frame-dependence of the dynamics of the scale factor \cite{rev-2-1}, as well as of the thermodynamical properties of black holes with non-constant curvature scalar \cite{rev-2-2}, have been already covered in the bibliography. These results are in line with our own point of view stating that, given that there does not really exists a underlying conformal symmetry in the motion equations of scalar-tensor theories, the different conformal frames represent, to all purposes, different theories with their own sets of measurable quantities.

The fine point in the case of the $f(R)$ theories in what regards to the physical equivalence of the different mathematically equivalent formulations, is that a new ``degree of freedom'' adds to the discussion: are physically equivalent the otherwise mathematically equivalent formulations \eqref{met-fdr-action}/\eqref{pal-fdr-action} and their corresponding JFBD counterparts \eqref{met-bd-action}/\eqref{pal-bd-action}? In \cite{ruf} the authors investigated whether the classical (mathematical) equivalence of $f(R)$ gravity and its formulation as scalar-tensor theory still holds at the quantum level. It was found that the equivalence is broken by off-shell quantum corrections, but recovered on-shell.


\section{Final remarks on the classification of scalar-tensor theories of gravity}\label{sect-discu}

Let us to discuss on a fine point in connection with the equations that define the measured Newton's constant in the case of Horndeski theories. Let us to illustrate the discussion with two Horndeski-type theories that have played an important role in cosmological applications and where the derivative couplings (and self-couplings) play an important part: the cubic galileon based on the action \eqref{ef-qbic-gal} and the theory with kinetic coupling to the Einstein's tensor that is based on the action \eqref{k-coup-action}. Since in both cases the function $G_4$ is a constant: $G_4=1/2$, according to \eqref{newton-c}, the measured Newton's constant in local (Solar System) experiments coincides with the one in general relativity: $G_\text{eff}=(8\pi)^{-1}M_\text{Pl}^{-2}$. I. e., local experiments can not differentiate between the cubic galileon, the kinetic coupling and general relativity theories. This is to be contrasted with the fact that, according to \eqref{qbic-class-eff-g} and to \eqref{k-coup-gn}, where the measured Newton's constant is derived in a cosmological setting, the cosmological observations allow to clearly differentiate the cubic galileon from the kinetic coupling theory and both latter theories from general relativity. This apparent inconsistency is due to the fact that the equation \eqref{newton-c} -- the same for \eqref{newton-c'} -- is derived within the PPN formalism and is not applicable to theories where the screening mechanisms play a role. This is the case for the cubic galileon and the kinetic coupling theories where the Vainshtein screening \cite{vainshtein} is the dominant effect for distances far below the Vainshtein radius.

The fact is that, even if assume that \eqref{newton-c} were a valid equation for the theories with the Vainshtein screening, if we compare equations \eqref{newton-c} -- or its equivalent \eqref{newton-c'} -- and \eqref{horn-eff-g}, it is seen that these can be matched only in the massless case. Yet it is not required that these matched at local scales since, in the vicinity of massive sources where the expression \eqref{newton-c'} is useful, below the Vainshtein radius $$r_V=\left(\frac{M}{8\pi M_\text{Pl}\Lambda^3}\right),$$ where $M$ is the mass of the source and the scale $\Lambda\sim M_\text{Pl}H_0^2$ ($H_0$ is the present value of the Hubble parameter), the non-linear contribution coming from second derivatives of the scalar field starts dominating, which results in that at $r\ll r_V$ the kinetic term of the scalar field decouples from the rest of the matter degrees of freedom through the Vainshtein mechanism \cite{vainshtein}. In this highly non-linear regime, the given Horndeski theory is well approached by general relativity. Notice that, the Vainshtein radius of the Sun $r_V^\text{Sun}\sim 10^{18}$m, while the size of the solar system $r_\text{SS}\sim 10^{12}$m, so that $$\frac{r_\text{SS}}{r_V^\text{Sun}}\sim 10^{-6}\ll 1.$$ This means that general relativity is a good approximation within the Solar system. Hence, the conclusion that Solar system experiments are not able to differentiate between theories where the Vainshtein screening arises and general relativity, is correct whether or not \eqref{newton-c} is valid. This means that, unlike cosmological observations, local experiments are `blind' to the derivative couplings.


\section{Conclusion}\label{sect-conclu}

In this paper we have proposed an unambiguous classification of the scalar-tensor theories of gravity. We have included the Horndeski-type of generalization that arises when derivative couplings are considered. In this case, thanks to the Vainshtein screening, the different theories are very well approached by general relativity in Solar system experiments, meanwhile in a cosmological framework these theories may be well differentiated. The unambiguous classification of the gravitational theories into metric scalar-tensor theories and non-metric general relativity with a scalar field non-minimally coupled to matter is possible only if the different conformal frames in which a given STT may be formulated were not physically equivalent. Although some times speculated, it is not guarantee that the quantum effects will break the conformal equivalence and thus spoil the classification we have discussed here, so that our results might still hold in the quantum regime.

Let us summarize the main achievements of this paper.

\begin{itemize}

\item We have shown that an incorrect classification of scalar-tensor theories, as in the case of the Saez-Ballester theory \eqref{saez-ballester-action}, may lead to misunderstanding and to the derivation of spurious physical consequences \cite{saez-ballester, socorro-2010, jamil-epjc-2012, sharif-prd-2014, rao-2017, aditya-2018, rao-2018, mishra-2019}. 

\item A criterion for an unambiguous classification of the STT-s has been proposed and several concrete examples have been worked out in order to show how this criterion works.

\item We have shown how the derivative coupling to the curvature, not evident at the Lagrangian level, arises in the cubic galileon model \eqref{qbic-class} as a consequence of first-order variations of the cubic galileon Lagrangian.

\item It is a fact that the expressions for the measured Newton's constant in Horndeski theories that are derived through the PPN formalism \eqref{newton-c} and through the cosmological perturbations \eqref{horn-eff-g} approach, do not coincide since the PPN formalism is not applicable to theories where screening mechanisms play a role. We have shown that Solar system experiments are blind to the derivative couplings thanks, precisely, to the Vainshtein screening given that the ratio $r_\text{SS}/r_V^\text{Sun}\sim 10^{-6}$, so that general relativity is a good approximation within the Solar system. Hence, within the Solar system we may use the expression \eqref{newton-c} with confidence despite of the importance the Vainshtein screening may carry for the given Horndeski theory.

\item We have discussed on the role the conformal transformations issue have on the unambiguous classification of the STT-s. In particular, if the different conformal frames in which a given scalar-tensor theory may be formulated were physically equivalent, then there were no point in classifying the different theories into scalar-tensor or general relativity with a scalar field either minimally or non-minimally coupled to the matter d.o.f. The same holds true for the classification of given theories of gravity into non-metric, metric, bi-metric, etc.

\end{itemize}

We want to underline that the discussion in the present paper was limited to standard (torsionless) scalar-tensor theories and to their generalizations belonging in the Horndeski class. Other known modifications of the gravitational theory as the teleparallel theories, 'beyond Horndeski' theories, the new gravitational scalar-tensor theories and the extended theories of gravity, were excluded from the present discussion. A proper classification of scalar-tensor theories with application to these modifications will be the subject of forthcoming work.


\section{Acknowledgments}

Useful comments by N Dimakis, S D Odintsov, S Ippocratis, I D Saltas, S Karamitsos and C F Steinwachs are acknowledged. The authors are grateful to SNI-CONACyT for continuous support of their research activity. RDA also acknowledges PRODEP for the postdoc grant under which part of this work was performed. The work of RGS was partially supported by SIP20190237, COFAA-IPN, and EDI-IPN grants.


\end{document}